\documentclass[english,twoside,a4paper,10pt]{article}

\usepackage[latin1]{inputenc}
\usepackage[T1]{fontenc}
\usepackage{dsfont}
\usepackage{amsmath}
\usepackage{amsfonts}
\usepackage{graphicx}
\usepackage{subfigure}
\usepackage{a4wide}
\usepackage{amssymb}
\usepackage{fancyhdr}
\usepackage{mathrsfs}
\usepackage[toc,page]{appendix}

\begin{document}

\title{\bf Dark Viscous Fluid coupled with Dark Matter and future singularity}
\author{ Lorenzo Sebastiani\footnote{E-mail address:
l.sebastiani@science.unitn.it
}\\
\\
\begin{small}
Dipartimento di Fisica, Universit\`a di Trento \end{small}\\
\begin{small}and Istituto Nazionale di Fisica Nucleare\end{small}\\ 
\begin{small}Gruppo Collegato di Trento, Italia
\end{small}\\
}
\date{}

\maketitle

\def\thesection{\Roman{section}}
\def\theequation{\Roman{section}.\arabic{equation}}

%%%%%%%%%%%%%%%%%%%%%%%%%%%%%%%%%%%%%%%%%%%%%%%%%%%%%%%%%%%%%%%%%%%%%%%%%%%%%%%%%%%%%%%%%%%%%%%%%%%%%%%%%%%%%%%%%%%%%%%%%%%%%%%%%%%

%%%%%%%%%%%%%%%%%%%%%
%  Abstract
%%%%%%%%%%%%%%%%%%%%%
\begin{abstract}

We study effects of viscous fluid coupled with dark matter in our universe.
We consider bulk viscosity in the cosmic fluid and we suppose the existence of a coupling between fluid and dark matter, in order to reproduce a stable de Sitter universe protected against future-time singularities. More general inhomogeneous fluids are studied related to future singularities. 

\end{abstract}
%%%%%%%%%%%%%%%%%%%%%

%----------------------------
%PACS
%----------------------------

%===========================================================================

%%%%%%%%%%%%%%%%%%%%%%%%%%%
%%%  Sec. I
%%%%%%%%%%%%%%%%%%%%%%%%%%%
\section{Introduction}

% The presence of viscous fluids in the cosmological scenario could bring the future universe evolution to finite-time singularities, in wich some physical quantities (like, for example, the scale factor $a$ or, more in general, the curvature) diverge.
% 
% 
\paragraph*{} Recent observational data imply that the current expansion of the universe
is accelerating\cite{accelerazione}. This is the so called Dark Energy issue.
There exist several descriptions of  this fact. Among them, the simplest one is the introduction of small positive 
 Cosmological Constant in the framework of General Relativity (GR), the so called $\Lambda$-CDM model, but in general also phantom/quintessence dark energy (DE) models with effective EoS parameter $\omega$ approximately equal to $-1$ are not excluded.  

Unfortunately, many of such DE-models  
bring the future universe evolution to finite-time singularity.
The classification of the (four) finite-time future singularities has been made
in Ref.\cite{problemisingolari}. Some of these types future singularities are softer than other and not all physical quantities (scale factor, effective energy density and pressure) necessarly diverge at this finite future time.

Note that singular solutions correspond to accelerated universe, and often appear as the final evolution of unstable de Sitter space. 

The presence of finite-time future singularities may cause serious problems in the black holes or stellar astrophysics\cite{Maeda}. Thus, it is of some interest to understand if any natural scenario exists in order to cure such singularities. 

In the recent work of Ref.\cite{Odintsov}, it has been considered the scenario of future singularity removal for coupled phantom energy with dark matter and it has been shown the existence of unstable de Sitter solution which may solve the coincidence problem but does not cure definitely the singularity. 

The purpose of present work is to generalize the results of Ref.\cite{Odintsov}, by considering the dark matter coupled with viscous fluid, as it is shown in Sec. II. In Sec. III we briefly discuss how the presence of viscous fluid could produce the singularities and several examples are given (for a review, see Ref.\cite{Sebastiani3}). Next, Sec. IV and Sec. V are devoted to reconstruct the standard cosmology and the stable de Sitter space avoiding the singularities for this kind of model. In Sec. VI, we explore the de Sitter solution in the presence of a generic class of coupled viscous fluids with non constant EoS parameter $\omega$ and we recover the Cardy-Verlinde formula, whereas in Sec. VII we consider several examples of inhomogeneous fluid and the singularity avoidance as a consequence of higher-order corrections of modified gravity.
Finally, conclusions are given in Sec.\ VIII.

%%% Unit %%%
We use units of $k_\mathrm{B} = c = \hbar = 1$ and denote the gravitational constant $G_{N}$. 
% denote the
% gravitational constant $8 \pi G_{N}$ by
% ${\kappa}^2 \equiv 8\pi/{M_{\mathrm{Pl}}}^2$
% with the Planck mass of $M_{\mathrm{Pl}} = G_{N}^{-1/2} =
% 1.2 \times 10^{19}$GeV.
%%%%%%%%%%%%
%%%%%

%%%%%%%%%%%%%%%%%%%%%%%%%%%%%%%%%%%%%%%%%%%%%%%%%%%%%%%%%%%%%%%%%%%%%%%%%%%%%%%%%%%%%%%%%%%%%%%%%%%%%%%%%%%%%%%%%%%%%%%%%%%%%%%%%%%
\setcounter{equation}{0}

\section{Formalism}

\paragraph*{} Let us recall several fundamental points about GR. The spatially-flat FRW space-time is described by the metric
\begin{equation}
ds^{2}=-dt^{2}+a^{2}(t)d\mathbf{x}^{2}\,,
\end{equation}
where $a(t)$ is the scale factor of the universe. The Hubble parameter is $H=\dot{a}/a$ (the point denotes the derivative with respect to the cosmic time $t$).

The FRW equations of motion (EOM) are:
\begin{equation}
\rho=\frac{3}{8\pi G_{N}}H^{2}\label{EOM1}\,,
\end{equation}
\begin{equation}
p=-\frac{1}{8 \pi G_{N}}\left(2\dot{H}+3H^2\right)\label{EOM2}\,,
\end{equation}
where $\rho$ and $p$ are, respectively, the total energy density and the total pressure of contents of the universe and describe ordinary matter, radiation, dark matter (DM) or more general fluids able to explain the accelerated expansion of the universe today.

From Eq.(\ref{EOM1}) and Eq.(\ref{EOM2}), we get the energy conservation law
\begin{equation}
\dot \rho +3H(\rho+p)=0\,.\label{ECL}
\end{equation}

It is well-know that ordinary matter and radiation are decoupled and separately satisfy the same form of energy conservation law, but it is not necessarily true for other kinds of energy.    
 
In what follows, we will consider the model where a viscous fluid and dark matter are coupled. Their energy conservation laws are given by:
\begin{equation}
\dot \rho_{F} +3H(\rho_{F}+p_{F})=-Q\rho_{F}\,,\label{ECL1}
\end{equation}
\begin{equation}
\dot \rho_{DM} +3H\rho_{DM}=Q\rho_{F}\,.\label{ECL2}
\end{equation}
Here, $Q$ is assumed to be a constant, $\rho_{DM}$ is the energy density of dark matter (the pressure of DM is equal to zero), whereas $\rho_{F}$ and $p_{F}$ are the energy density and pressure of viscous fluid. 

We assume the fluid equation of state (EoS) in the form\cite{Alessia, Motta}:
\begin{equation}
p_{F}=\omega(\rho_{F})\rho_{F}-3 H\zeta(\rho_{F})\,,\label{eq.state}
\end{equation}
$3H$ being the four-velocity of the cosmic fluid. The thermodynamical variable $\omega(\rho_{F})$ is an arbitrary function of the density $\rho_{F}$. $\zeta(\rho_{F})$ is the bulk viscosity and in general it depends on $\rho_{F}$. On thermodynamical grounds, in order to have the positive sign of the entropy change in an irreversible process, $\zeta(\rho_{F})$ has to be a positive quantity, so we require $\zeta(\rho_{F})>0$. Note that such viscous fluid represents special case of more general
inhomogeneous fluid introduced in Ref.\cite{inEOS}. 

By taking into account Eq.(\ref{eq.state}), we can rewrite Eq.(\ref{ECL1}) in the following form: 
\begin{equation}
\dot{\rho_{F}}+3H\rho_{F}(1+\omega(\rho_{F}))+Q\rho_{F}=9H^{2}\zeta(\rho_{F}) \label{conservationlaw}\,,
\end{equation}
and, in general, one has $\zeta(\rho_{F})$ as a function of $H$, $\zeta(H)$.

The effective EoS parameter $\omega_{eff}$ of the universe is
\begin{equation}
\omega_{eff}=\frac{p}{\rho}=-1-\frac{2\dot{H}}{3H^{2}}\,, \label{omegaeffettivo}
\end{equation}
where the EOM have been used. If $\omega_{eff}$ is less than $-1/3$, $\ddot a>0$ and the universe expands in accelerated way.

We want to study possible final evolution of the universe due to the presence of dark matter coupled with viscous fluid.

The non viscous case ($\zeta(H)=0$) has been discussed in Ref.\cite{Odintsov}, so in this work we suppose $\zeta(H)\neq 0$.

% \begin{equation}
%  R= 6\left(2H^{2}+\dot{H}\right)\label{R}\,,
% \end{equation}
% \begin{equation}
%  G=24 H^{2}\left(H^{2}+\dot{H}\right)\label{G}\,.
% \end{equation}

% For the stress-energy tensor $T_{\mu\nu}$, one has :
% \begin{equation}
% T_{\mu\nu}=\rho u_{\mu}u_{\nu}+\left(\omega(\rho)\rho-3H\zeta(\rho)\right)(g_{\mu\nu}+u_{\mu}u_{\nu})\,. 
% \end{equation}

% Note that the effective energy density of the universe is $\rho_{eff}=\rho_{G}+\rho$ and the effective pressure is $p_{eff}=p_{G}+p$, whereas for the effective equation of state, we obtain:
% \begin{equation}
% \omega_{eff}\equiv \frac{p_{eff}}{\rho_{eff}}= -1-\frac{2\dot{H}}{3H^{2}}\,. \label{Lorenzobis}
% \end{equation}

\setcounter{equation}{0}

\section{Singular future universe}

\paragraph*{} 

The presence of viscous fluids in the cosmological scenario could bring the future universe evolution to finite-time singularities, in which the Hubble parameter is expressed as:
\begin{equation}
H=\frac{h}{(t_{0}-t)^{\beta}}\label{Hsingular}\,,
\end{equation}
where $h$ and $t_{0}$ are positive constants and $t<t_{0}$ because it should be for expanding universe. $\beta$ is a positive constant or a negative non-integer number, so that, when $t$ is close to $t_{0}$, $H$ or some derivative of $H$ and therefore the curvature become singular. 

Note that the effect of viscous fluids to Big Rip singularity ($\beta=1$ in Eq.(\ref{Hsingular})) has been
discussed also in Ref.\cite{Brevik}. 

Such choice of Hubble parameter corresponds to accelerated universe, because if Eq.(\ref{Hsingular}) is a solution of the EOM (\ref{EOM1})-(\ref{EOM2}), it is easy to see that the strong energy condition ($\rho+3p\geqslant 0$) is always violated when $\beta>0$, or is violated for small value of $t$ when $\beta<0$. It means that $\omega_{eff}$ of Eq.(\ref{omegaeffettivo}) is less than $-1/3$ near the singularity.

The finite-time future singularities can be classified in the following way\cite{problemisingolari}: 
\begin{itemize}
\item Type I (Big Rip): for $t\rightarrow t_{0}$, $a(t)\rightarrow\infty$,
$\rho\rightarrow\infty$ and
$|p|\rightarrow\infty$. The case in which
$\rho$ and $p$ are finite at $t_{0}$ is also
included.
It corresponds to $\beta=1$ and $\beta>1$.
\item Type II (sudden):
for $t\rightarrow t_{0}$, $a(t)\rightarrow a_{0}$,
$\rho\rightarrow\rho_{0}$ and $|p|
\rightarrow\infty$.
It corresponds to $-1<\beta<0$.
\item Type III: for $t\rightarrow t_{0}$, $a(t)\rightarrow a_{0}$,
$\rho\rightarrow\infty$ and
$|p|\rightarrow\infty$.
It corresponds to $0<\beta<1$.
\item Type IV: for $t\rightarrow t_{0}$, $a(t)\rightarrow a_{0}$,
$\rho\rightarrow 0$, $|p|
\rightarrow 0$
and higher derivatives of $H$ diverge.
The case in which $\rho$ and/or $p$ tend to finite values is
also included. It corresponds to
$\beta<-1$ but $\beta$ is not any integer number.
\end{itemize}
Many of DE-models may lead to one of such future singularities in the universe evolution. The simplest case is the phantom perfect fluid (with equation of state $p_{DE}=\omega\rho_{DE}$ and $\omega<-1$). For this kind of fluid, the EOM (\ref{EOM1})-(\ref{EOM2}) admit the solution
\begin{equation}
H=-\frac{2}{3(1+\omega)}\frac{1}{(t_{0}-t)},
\end{equation}
which corresponds to Big Rip singularity for $\beta=1$ and $h=-2/3(1+\omega)$ in Eq.(\ref{Hsingular}).

Then, all singularities could appear if we consider the contribute of a non-zero bulk viscosity into the fluid equation of state (\ref{eq.state}).  

Let us consider several simple examples\cite{Sebastiani3}, when the thermodinamical parameter of viscous fluid $\omega$ is a constant, and the fluid is dominant and weakly coupled with dark matter ($Q\simeq 0$), so that its energy conservation law leads:
\begin{equation}
\dot{\rho_{F}}+3H\rho_{F}(1+\omega)\simeq 9H^{2}\zeta(\rho_{F}) \label{Hallo}\,.
\end{equation}

Suppose to have bulk viscosity proportional to $H$
\begin{equation}
\zeta(H)=\zeta_{0}H\,,\label{esempi}
\end{equation}
where $\zeta_{0}$ is a positive constant.

%The Big Rip singularity could appear if $\alpha=0,1$. 

The solution of Eq.(\ref{Hallo}) on the Big Rip singularity ($H=h/(t_{0}-t)$) is:
\begin{equation}
\rho_{F}\simeq\frac{9 h^{3}\zeta_{0}}{(2+3 h(1+\omega))(t_{0}-t)^{2}}\,,\label{prova}
\end{equation}
that solves the EOM when
\begin{equation}
 H=\frac{2}{(24\pi G_{N}\zeta_{0}-3(1+\omega))}\frac{1}{(t_{0}-t)}\,,
\end{equation}
under the requirement $24\pi G_{N}\zeta_{0}-3(1+\omega)>0$.

For the same model, also the so called Type I singularity ($\beta>1$ in Eq.(\ref{Hsingular})) could appear. In this case an asymptotic solution of Eq.(\ref{Hallo}) is:
\begin{equation}
\rho_{F}\simeq \frac{3h^{2}\zeta_{0}}{(1+\omega)(t_{0}-t)^{2\beta}}\,,
\end{equation}
which solves the EOM for $\omega>-1$.
 
In the next sections, we consider the model where viscous fluid couples with dark matter and we investigate if there exist other realistic possible final scenarios for the universe.

\setcounter{equation}{0}

\section{Coupling of viscous fluid with dark matter}

\paragraph*{} 

%We now consider model where viscous fluid couples with DM. 

Suppose to have $\omega$ constant for the fluid and bulk viscosity in the form
\begin{equation}
 \zeta=\zeta_{0}H^\alpha\,,
\end{equation}
where $\alpha$ is a real number. In this case, the general solution of Eq.(\ref{conservationlaw}) is
\begin{equation}
\rho_{F}=\rho_{0F}\frac{e^{-Q t-3\omega \log a(t)}}{a(t)^3}+\frac{9\zeta_{0} e^{-Q t-3\omega \log a(t)}}{a(t)^3}\int^{t} dt' e^{Q t'+3\omega \log a(t')}a(t')\dot{a}(t')^2\left(\frac{\dot{a}(t')}{a(t')}\right)^\alpha\,,\label{uno}
\end{equation}
where $\rho_{0F}$ is, as usually, a positive constant of integration.

One possible solution is the de Sitter space, where $H=H_{0}$ is a constant. One may identify the Hubble parameter $H_{0}$ with the present value of accelerated universe, $H_{0}\simeq 10^{-33}\text{ eV}$ in order to reconstruct the standard cosmology. In this case, Eq.(\ref{uno}) can be solved as
\begin{equation}
\rho_{F}=\rho_{0F}e^{-t(Q+3H_{0}(1+\omega))}+\frac{9H_{0}^{\alpha+2} \zeta_{0}}{(Q+3H_{0}(1+\omega))}\,.\label{due}
\end{equation}

It follows the solution of Eq.(\ref{ECL2}) for dark matter
\begin{equation}
\rho_{DM}=\rho_{0DM}e^{-3H_{0}t}-\rho_{0F}\frac{Q}{Q+3H_{0}\omega}e^{-t(Q+3H_{0}(1+\omega))}+\frac{3H_{0}^{\alpha+1}Q\zeta_{0}}{(Q+3H_{0}(1+\omega))}\,,\label{tre}
\end{equation}
where $\rho_{0DM}$ is a positive constant. It is easy to see that, if $\zeta_{0}\neq 0$, the EOM (\ref{EOM1})-(\ref{EOM2}) are satisfied only if $\rho_{0F}=\rho_{0DM}=0$. Therefore, we note that, if the de Sitter solution is an attractor and is able to describe our universe today, we can require
\begin{equation}
\frac{\rho_{DM}}{\rho_{F}}=\frac{Q}{3H_{0}}\sim \frac{1}{3}\,,
\end{equation}
and the coincidence problem is solved by setting
\begin{equation}
Q=H_{0}\,.\label{Q} 
\end{equation}

Now, the ratio of DM and fluid is approximately $1/3$, almost independent from initial conditions. 

By using Eq.(\ref{EOM1})-(\ref{EOM2}), one has the relation between $\omega$ and $\zeta_{0}$ necessary to describe the de Sitter space: 
\begin{equation}
 \omega=-\frac{4}{3}+\frac{4}{3}H_{0}^{\alpha-1}\zeta_{0}(8\pi G_{N})\,.\label{omega}
\end{equation}
Here, Eq.(\ref{Q}) has been used. With respect to the case of viscosity equal to zero (see Ref.\cite{Odintsov}), where, in order to solve the coincidence problem $\omega=-4/3$, now the generic condition on $\omega$ is $\omega>-4/3$, and $\rho_{F}$ of Eq.(\ref{due}) is correctly positive.

For example, a DE-fluid with $\omega=-1$ admits an exactly de Sitter solution for $H=H_{0}$ if its bulk viscosity is
\begin{equation}
\zeta=\frac{H^{\alpha}}{32\pi G_{N}H_{0}^{\alpha-1}}\,, 
\end{equation}
and the coupling constant with DM is $Q=H_{0}$.

\setcounter{equation}{0}
\section{Cosmology}

\paragraph*{}During the matter dominated era, dark matter has to drive the expansion of the universe. In this case, $H_0=Q<<H$. We suppose $\rho_F<<\rho_{DM}$, so that Eq.(\ref{ECL2}) and Eq.(\ref{conservationlaw}) result decoupled:
\begin{equation}
\dot{\rho_{F}}+3H\rho_{F}(1+\omega)\simeq 9H^{\alpha+2}\zeta_0 \label{ECL1bis}\,,
\end{equation}
\begin{equation}
 \dot \rho_{DM} +3H\rho_{DM}\simeq 0\,.\label{ECL2bis}
\end{equation}

The scale factor behaves as 
\begin{equation}
a(t)=a_{0}t^{2/3}\,, \label{a}
\end{equation}
where $a_{0}$ is a positive constant. In this case, the solution of Eq.(\ref{ECL1bis}) is
\begin{equation}
\rho_{F}\simeq 9\zeta_{0}\left(\frac{2}{3}\right)^{\alpha+2}\frac{1}{(1+2\omega-\alpha)}\frac{1}{t^{\alpha+1}}\,,
\end{equation}
where we have omitted the homogeneus part. Since the solution of Eq.(\ref{ECL2bis}) is
\begin{equation}
\rho_{DM}\simeq \frac{\rho_{0DM}}{a_{0}^{3} t^2}\,,
\end{equation}
the ratio between energy density of fluid and dark matter when $H>>H_{0}$, behaves as
\begin{equation}
\frac{\rho_{F}}{\rho_{DM}}\simeq \frac{1}{\rho_{0DM}(8\pi G_{N})}\left(\frac{H_{0}}{H}\right)^{1-\alpha}\,,
\end{equation}
where Eq.(\ref{omega}) has been used.

It means that, if $\alpha<1$, the energy density of fluid increases with respect to the dark matter and the universe exits from matter era when the Hubble rate $H$ is close to $H_{0}$. 

Note that the condition $\alpha<1$ is not strictly necessary. When $H$ becomes close to $H_0$, the first order approximation of Eq.(\ref{uno}) on the solution (\ref{a}) becomes
\begin{equation}
\rho_{F}\simeq 9\zeta_{0}\left(\frac{2}{3}\right)^{\alpha+2}\frac{1}{(1+2\omega-n)}\left(\frac{1}{t^{\alpha+1}}-\frac{1}{(2+2\omega-n)}\frac{H_{0}}{t^{\alpha}}\right)+\mathcal{O}(H_{0}^2)\,,
\end{equation}
and so on for the successive orders of $H_{0}$: the term of $\rho_F$ proportional to $H_{0}^{n'}$ behaves as $1/t^{\alpha+1-n'}$, and in any case there exist a value of $H$ close to $H_{0}$ for which the fluid energy density improves, the EOM become inconsistent with respect to the solution (\ref{a}), and the universe exits from matter dominated era. 

A possible final scenario is the de Sitter universe described in the previous paragraph (we assume that the first term of Eq.(\ref{tre}) is negligible on the de Sitter solution).

In order to investigate if the de Sitter solution is an attractor or not, we consider the perturbation as 
\begin{equation}
H(t)=H_{0}+\Delta(t)\,.\label{perturbazione}
\end{equation}
Here, $\Delta(t)$ is assumed to be small. The second EOM (\ref{EOM2}) gives  
\begin{equation}
2\dot{\Delta}(t)+6H_{0}\Delta(t)\simeq 3H_{0}(\alpha+1)\Delta(t)\,,
\end{equation}
where we have used Eq.(\ref{due}) and Eq.(\ref{omega}). By assuming $\Delta(t)=e^{\lambda t}$, we find
\begin{equation}
 \lambda+3H_{0}-\frac{3}{2}H_{0}(\alpha+1)\simeq 0\,,
\end{equation}
that is
\begin{equation}
 \lambda\simeq \frac{3}{2}H_{0}(\alpha-1)\,.
\end{equation}

Then, if $\alpha<1$, the de Sitter solution is stable and the coupling of viscous fluid and dark matter at last generates a stable accelerated universe with a constant rate of DM and fluid. 

If $\alpha>1$, the de Sitter solution is not stable and other future scenarios are possible. For example, viscous fluid could generate a future singular solution and, if the singularity corresponds to the stable solution, the universe could approach to de Sitter space, but will finally evolve to such singularity. 

\setcounter{equation}{0}
\section{Viscous fluid with non constant omega and the de Sitter universe}

\paragraph*{}Let us consider a more general case, when the thermodinamical parameter $\omega$ of viscous fluid is not a constant, but it depends on the energy denisty $\rho_{F}$, so that $\omega=\omega(\rho_{F})$. A simple example is:
\begin{equation}
 \omega(\rho_{F})=A_{0}\rho_{F}^{\gamma-1}-1\,,\label{hallo}
\end{equation}
where $A_{0}$ and $\gamma$ are constant parameters. The energy conservation law of viscous fluid becomes
\begin{equation}
\dot\rho_{F}+3H A_{0}\rho_{F}^{\gamma}+Q\rho_{F}=9\zeta_{0}H^{\alpha+2}\,. 
\end{equation}
Here, we suppose the bulk viscosity proportional to $H^{\alpha}$, $\zeta=\zeta_{0}H^{\alpha}$. If we assume $\gamma>>1$, on the de Sitter solution $H=H_{0}$, we obtain 
\begin{equation}
\rho_{F}\simeq \left(\frac{3\zeta_{0}H_{0}^{\alpha+1}}{A_{0}}\right)^\frac{1}{\gamma}\,.\label{duebis}
\end{equation}

By using Eq.(\ref{ECL2}), the energy density of dark matter reads
\begin{equation}
 \rho_{DM}\simeq\frac{Q}{3H_{0}}\rho_{F}\,,
\end{equation}
and in order to solve the coincidence problem we have to require $Q=H_{0}$. 

From the EOM (\ref{EOM1})-(\ref{EOM2}), by assuming the fluid drives the accelerated expansion of the universe, it follows 
\begin{equation}
A_{0}\simeq 3\zeta_{0}H_{0}^{\alpha+1}\left(\frac{8\pi G_{N}}{3H_{0}^{2}}\right)^\gamma\,,\label{omegabis}
\end{equation}
and by using Eq.(\ref{hallo}), one has
\begin{equation}
\omega\simeq -1+H_{0}^{\alpha-1}\zeta_{0}(8\pi G_{N})\,,\label{sturmtruppen}
\end{equation}
that is an approximation of Eq.(\ref{omega}).

In order to investigate if the de Sitter solution is an attractor or not, we consider the perturbation as in Eq.(\ref{perturbazione}). The second EOM (\ref{EOM2}) gives  
\begin{equation}
2\dot{\Delta}(t)+6H_{0}\Delta(t)\simeq H_{0}\left(\frac{\alpha+1}{\gamma}\right)\Delta(t)\,,
\end{equation}
where we have used Eq.(\ref{duebis}) and Eq.(\ref{omegabis}). By assuming $\Delta(t)=e^{\lambda t}$, we find
\begin{equation}
 \lambda+3H_{0}-\frac{1}{2}H_{0}\left(\frac{\alpha+1}{\gamma}\right)\simeq0\,,
\end{equation}
that is
\begin{equation}
 \lambda\simeq H_{0}\left(\frac{1}{2}\left(\frac{\alpha+1}{\gamma}\right)-3\right)\,.
\end{equation}
Then, if $(\alpha+1)/\gamma<6$, the de Sitter solution is stable.\\

It could be of some interests the possibility to recover the Cardy-Verline (CV) formula for this kind of fluid on closed universe (with spatial curvature $k=1$). A study of the universality of CV formula for inhomogeneous fluids has been presented in Ref.\cite{CVF} (specifically for viscous fluids see Ref.\cite{B-O}). For a perfect fluid with $\omega$ constant, the entropy $S$ of the closed universe is written as a function of the total energy $E$ and the Casimir energy $E_{C}$ (for a review, see   Ref.\cite{genericCVFbis}, \cite{genericCVF}): 
\begin{equation}
 S=\left(\frac{2\pi a^{3\omega}}{\sqrt{\alpha\beta}}\sqrt{E_{C}(2E-E_{C})}\right)^{\frac{3}{3(1+\omega)-1}}\,,\label{CV}
\end{equation}
where $a$ is, as usually, the scale factor of the universe and $\alpha$ and $\beta$ are undetermined constants.

We assume an EoS in the form
\begin{equation}
p=A_{0}\rho_{F}^{\gamma}-\rho_{F}-3\zeta_{0}\,, 
\end{equation}
which is the case as above with bulk viscosity $\zeta=\zeta_{0}/H$. The energy conservation law leads
\begin{equation}
\rho'_{F}+\frac{3A_{0}\rho_{F}^{\gamma}}{a}=\frac{9\zeta_{0}}{a}\,, \label{tatatam}
\end{equation}
where the prime over $\rho_{F}$ denotes derivative with respect to the scale factor $a$. A simple solution of this equation is
\begin{equation}
\rho_{F}=\left(\frac{3\zeta_0}{A_{0}}\right)^{\frac{1}{\gamma}}\,. 
\end{equation}
The total energy inside the comoving volume $V$ is $E=\rho_F V$, so that $E\propto a^{3}$. For closed universe, we can write the energy as the sum of an extensive part $E_{E}$ and a subextensive part $E_{C}$, called the Casimir energy, and it takes the form\cite{genericCVF}:
\begin{equation}
E(S,V)=E_{E}(S,V)+\frac{1}{2}E_{C}(S,V)\,. \label{totalE}
\end{equation}

The extensive and subextensive parts of total energy, under a rescaling of entropy and volume, transform as
\begin{equation}
 E_{E}(\lambda S, \lambda V)=\lambda E_{E}(S,V)\,,\phantom{space}E_{C}(\lambda S, \lambda V)=\lambda^{1/3}E_{C}(S,V)\,,\label{rescaling}
\end{equation}
where $\lambda$ is a constant.

The FRW universe expands adiabatically, $dS=0$, and the products $E_{C}a^{-3}$ and $E_{E}a^{-3}$ should be independent on the volume $V$. Then, by using the rescaling properties of (\ref{rescaling}), we can write $E_{E}$ and $E_{C}$ as functions of the entropy
\begin{equation}
E_{E}=\frac{\alpha a^3}{4\pi }\,,\phantom{space}E_{C}=\frac{\beta a^3}{2\pi }S^{-2/3}\,. 
\end{equation}
By using Eq.(\ref{totalE}), we obtain:
\begin{equation}
  S=\left(\frac{2\pi a^{-3}}{\sqrt{\alpha\beta}}\sqrt{E_{C}(2E-E_{C})}\right)^{-3}\,.
\end{equation}
This expression is very similar to Eq.(\ref{CV}). The CV formula is recovered for $\omega=1$, but in general only for some special choices or solutions on closed universe, we have an identification between the FRW equation of this kind of fluid and the CV formula.

\setcounter{equation}{0}
\section{Inhomogeneous dark fluid and singularity avoidance}

\paragraph*{} In this paragraph we will consider inhomogeneous dark fluid, whose equation of state is given by\cite{inEOS} (see also Ref.\cite{aggiunta}):
\begin{equation}
p_{F}=\omega(\rho_{F})\rho_{F}+B(\rho_{F},a(t),H, \dot{H}...)\,,\label{start}
\end{equation}
where $B$ is a function of $\rho_{F}$, $a(t)$, $H$ and the derivatives of $H$. The motivation for this general form of time-dependent bulk viscosity comes from the modification of gravity. 

We want to consider several examples, where specific models producing future-time singularity are cured by coupled of fluid with dark matter or by adding viscosity term into the EoS. 

Let us start with the following simple EoS:
\begin{equation}
p_{F}=-\rho_F+f(\rho_{F})\,,
\end{equation}
where $f(\rho_{F})$ is a function of the fluid energy density $\rho_{F}$. In Ref.\cite{inEOS} is reconstructed the form of $f(\rho_{F})$ in order to have the singularity when the fluid energy conservation law is expressed as
\begin{equation}
\dot{\rho}_F+3H f(\rho_F)=0\,.
\end{equation}

By using the EOM (\ref{EOM1})-(\ref{EOM2}), we obtain the following solution of the scale factor
\begin{equation}
a(t)=a_{0}\left(\frac{t}{t_{0}-t}\right)^n\,, \label{BigRip}
\end{equation}
when the form of $f(\rho_{F})$ is
\begin{equation}
f(\rho_F)=\pm\frac{2\rho_F}{3n}\left(1-\frac{4n}{t_{0}}\left(\frac{3}{8\pi G_{N}\rho_F}\right)^{\frac{1}{2}}\right)^{\frac{1}{2}}\,. \label{f}
\end{equation}
In Eq.(\ref{BigRip}), $n$ is a positive constant and $t<t_{0}$, so that the Hubble parameter diverges in a finite time ($t\rightarrow t_{0}$) as in the Big Rip:
\begin{equation}
 H=\frac{nt_{0}}{t(t_{0}-t)}\,.\label{bip}
\end{equation}
Therefore, when $t<<t_{0}$, $H(t)$ evolves as $n/t$, which means that the effective EoS parameter $\omega_{eff}$ of Eq.(\ref{omegaeffettivo}) is given by $\omega_{eff}\simeq -1+2/(3n)>-1$. On the other hand, when $t\sim t_{0}$, $H(t)\simeq n/(t_0-t)$ and $\omega_{eff}\simeq -1-2/(3n)<-1$. The transition between the region $\omega_{eff}>-1$ and $\omega_{eff}<-1$ occours when $t\simeq t_{0}/2$.

Consider now the coupling of this fluid with dark matter as in Eq.(\ref{ECL1})-(\ref{ECL2}). Since $\omega$ parameter of DM is equal to zero and violates the condition $\omega<-1$ for which the Big Rip singularity appears, only the fluid could drive the universe to such singularity, so we assume that  singular solution of Eq.(\ref{bip}) already exist and 
\begin{equation}
\rho_{F}\simeq\frac{3 H^2}{8\pi G_{N}} =\frac{3n^2 t_{0}^2}{8\pi G_{N}t^2(t_{0}-t)^2}\,. \label{eins}
\end{equation}

We note that during the crossing of phantom barrier ($\omega_{eff}=-1$), $H(t)\simeq 2n/t$ and the solution of Eq.(\ref{ECL2}) is:
\begin{equation}
\rho_{DM}\simeq \frac{12 Q n^2}{8\pi G_N(6n-1)t}\,, \label{zwei}
\end{equation}
evaluated in $t=t_{0}/2$. 

We observe that, if
\begin{equation}
\frac{Q}{(6n-1)}>\frac{2}{t_{0}}\,, 
\end{equation}
the hypothesis that fluid energy density of Eq.(\ref{eins}) is dominant with respect to the DM energy density of Eq.(\ref{zwei}) when $t=t_0/2$ is violated. For large value of coupling constant $Q$, the universe does not cross phantom barrier and is protected against Big Rip future singularity occouring when $t=t_{0}$. Note that $t_{0}$ is determinated by explicit form of $f(\rho)$ in Eq.(\ref{f}).\\

Let us consider the possibility to obtain the inhomogeneous EoS from the modified gravity. The following action is considered:
\begin{equation}
S=\int d^{4}x\sqrt{-g}\left(\frac{1}{16\pi G_{N}}(R+f(R, G))\right)\,,
\end{equation}
where $g$ is the determinant of the metric tensor and $f(R, G)$ is a function of the Ricci scalar $R$ and the Gauss-Bonnet invariant $G$. By using the variational principle, we can write the FRW equations 
\begin{equation}
\tilde{\rho}=\frac{3}{8\pi G_{N}}H^{2}\,,
\end{equation}
\begin{equation}
\tilde{p}=-\frac{1}{8 \pi G_{N}}\left(2\dot{H}+3H^2\right)\,.
\end{equation}
Here, modified gravity has been included in the energy density $\tilde{\rho}$ and the pressure $\tilde{p}$ of inhomogeneous fluid so that:
% The part of modified gravity is formally included into the modified energy density $\rho_{G}$ and the modified pressure $p_{G}$ as follows:
\begin{equation}
\tilde{\rho}=-\frac{1}{16\pi G_{N}}\Bigl\{24H^{3}\dot{f}'_{G}+6H^{2}f'_{R}+6H\dot{f}'_{R}+(f-Rf'_{R}-Gf'_{G})\Bigr\}\label{rhoG}\,,
\end{equation}
\begin{eqnarray}
\tilde{p}&=&\frac{1}{16\pi G_{N}}\Bigl\{8H^{2}\ddot{f}'_{G}+2\ddot{f}'_{R}+4H\dot{f}'_{R}+16H\dot{f}'_{G}(\dot{H}+H^{2})+f'_{R}(4\dot{H}+6H^{2})+\nonumber \\ \nonumber \\ & & \phantom{spacespa}(f-R\dot{f}'_{R}-Gf'_{G})\Bigr\}\label{pG}\,.  
\end{eqnarray}
The point denote the derivative with respect to the cosmic time, and we have used the following expressions:
\begin{equation}
f'_{R}=\frac{\partial f(R,G)}{\partial R}\,,\phantom{spacespacespace}f'_{G}=\frac{\partial f(R,G)}{\partial G}\,. 
\end{equation}
Thus, it follows
\begin{equation}
\tilde{p}=\omega\tilde{\rho}+B(H,\dot{H}...)\,, 
\end{equation}
where
\begin{eqnarray}
\hspace{-5mm}
B(H,\dot{H}...) &=&
\frac{1}{16\pi G_{N}} \biggl\{ (1+w)(f
-R f'_{R}-G f'_{G})
\nonumber \\
\hspace{-5mm}
&&
+\left(R+f'_{R}\right)
\left[6H^2(1+w)+4\dot{H}\right]
\nonumber \\
\hspace{-5mm}
& &
+ H{\dot f}'_{R}(4+6w)
+8H{\dot f}'_{G}\left[2\dot{H}+ H^{2}(2+3w)\right]
+2{\ddot f}'_{R}+8 H^{2}{\ddot f}'_{G} \biggr\}\,,
\label{Feldwebel}
\end{eqnarray}
and we recover the EoS in the form of Eq.(\ref{start}) by writing 
\begin{equation}
R=6\left(2H^2+\dot{H}\right)\,,\phantom{spacespace}G=24H^2\left(H^2+\dot{H}\right)\,. 
\end{equation}

In Ref.\cite{mg} are shown several models of $f(R,G)$-modified gravity producing future-time singularities. For example, in the model $f(R, G)=-\alpha G/R$, where $\alpha$ is a positive constant, could appear the Type I singularity, whereas in the model $f(R,G)=\alpha R^{\gamma}$, where $\alpha$ and $\gamma$ are constants, could appear Types II, III or IV singularities.

It is well know that, in order to cure the singularity problem in $f(R,G)$-gravity, it is possible to use some power function of $R$, via scenario first
suggested in Ref.\cite{Abdalla} (for more details, see Ref.\cite{mg2}), or power function of $G$. As a result, term $R^{m}$ with $m>1$ cures Types I, II and III singularities. Moreover, if $m< 2$, the Type IV singularities are cured. The same happens with the terms $G^{m}$: $m>1/2$ cures the Type I, II and III singularities, $m\leqslant 0$ the Type IV.  
The Type I singularities can be avoided also by using term $G^m R^n$, where $m$ and $n$ are positive numbers, and so on. Scenario of singularity avoidance for realistic modified
gravity is discussed in Ref.\cite{mg2}, \cite{realisticmodel}.

This kind of terms could be seen as quantum effects in large curvature regime, or could be included in the viscosity part $B(H, \dot{H}...)$ in the equivalent description of inhomogeneous fluid.

\setcounter{equation}{0}
\section{Conclusion}

\paragraph*{} In the present paper, we have investigated the possible final scenarios for the universe, in presence of viscous fluid coupled with dark matter. 

In principle, viscous fluids could generate any of the known finite-time future singularities, which emerge from accelerated universe. If the singularity corresponds to stable solution, the universe could finally evolve to such singularity. A way to prevent this possibility is to consider the effects of viscosity on the de Sitter solution descibing the universe today.  

As a result, the presence of bulk viscosity gives rise to a large class of models able to explain the current acceleration by producing a stable de Sitter solution protected against singularities. With coupling fluid and DM, the coincidence problem could be solved, as first it has been demonstrated in Ref.\cite{Odintsov}, and the viscosity allows to have the stability of De Sitter solution.

As final remark, we note that inhomogeneous EoS DE/modified gravity may cure singularities including corrections in high curvature regime. 

%%%%%%%%%%%%%%%%%%%%%%%%
%%%  Acknowledgments
%%%%%%%%%%%%%%%%%%%%%%%%
\section*{Acknowledgments}
\paragraph*{}L.S. thanks 
%Professor Sergei Odintsov and
Professor Sergio Zerbini for comments and valuable suggestions on this work. The work is supported in part by INFN (Trento)-CSIC (Barcelona).
% Casimir Effect.
% %%% bozza
% K.B. and S.D.O. thank Professor Shin'ichi Nojiri for his
% collaboration in the previous work~\cite{Odintsov}.
% %%%
% The work is supported in part by
% the National Science Council of R.O.C. under
% Grant \#s: NSC-95-2112-M-007-059-MY3 and NSC-98-2112-M-007-008-MY3 and
% National Tsing Hua University under the Boost Program and Grant \#:
% 97N2309F1 (K.B.);
% MEC (Spain) project FIS2006-02842 and AGAUR (Catalonia) 2009SGR-994, by
% JSPS Visitor Program (Japan) and by LRSS project
% N.2553.2008.2 (S.D.O.).

\end{document}